\documentstyle[twoside,fleqn,espcrc2,epsfig]{article}

\def\beq{\begin{equation}}
\def\eeq{\end{equation}}
\def\barr{\begin{eqnarray}}
\def\earr{\end{eqnarray}}
\def\lsim{\raise0.3ex\hbox{$\;<$\kern-0.75em\raise-1.1ex\hbox{$\sim\;$}}}
\def\gsim{\raise0.3ex\hbox{$\;>$\kern-0.75em\raise-1.1ex\hbox{$\sim\;$}}}
\def\lg{\raise0.3ex\hbox{$\;>$\kern-0.75em\raise-1.1ex\hbox{$<\;$}}}
\def\equiapp{\raise0.3ex\hbox{$\;\sim$\kern-0.75em\raise-1.1ex\hbox{$=\;$}}}

\newcommand{\AmS}{{\protect\the\textfont2
  A\kern-.1667em\lower.5ex\hbox{M}\kern-.125emS}}

\title{Resolving Ambiguities in the Neutrino Mass-Flavour Spectrum from
Supernova Neutrinos}

\author{Amol S. Dighe 
  \address{Div. TH, CERN, CH-1211 Geneva 23, Switzerland.}
  }

\begin{document}

\begin{abstract}
We analyze the neutrino conversions inside a supernova
in the 3$\nu$ mixing scheme, and 
their effects on the neutrino spectra observed at the earth.
We find that the observations of the energy spectra 
of neutrinos from a future galactic supernova 
may enable us to identify the solar neutrino solution, 
to determine the sign of $\Delta m^2_{32}$,
and to probe the mixing matrix element 
$|U_{e3}|^2$ to values as low as $10^{-4}-10^{-3}$.

\end{abstract}

\maketitle

\section{Introduction}

In the solutions of the solar and atmospheric
neutrino anomalies through the oscillations
between the three active neutrino species, 
three ambiguities remain to be resolved:
(i) the  solution of the solar neutrino problem --
SMA, LMA or VO,
(ii) the type of mass hierarchy --
normal ($m_3 > m_1, m_2$) or inverted ($m_3 < m_1, m_2)$, and
(iii) the  value of $|U_{e3}|^2$.
We show how some of these ambiguities may be
resolved from the observations of the
neutrino spectra from a Type II supernova.

Only the main results are summarized in this talk. For
the detailed arguments and derivations, the reader is referred to
\cite{ds}.

\section{Neutrino Conversions}

The neutrino transitions between different matter eigenstates
inside the supernova take place mainly in the resonance
regions $H$ and $L$, which are characterized by 
$(\Delta m^2_{atm},4|U_{e3}|^2)$ and 
$(\Delta m^2_\odot, \sin^2 2\theta_\odot)$ respectively.
Due to the $\Delta m^2$-hierarchy ($\Delta m^2_{atm} \gg
\Delta m^2_\odot$), the dynamics in 
each of the two resonance layers
can be considered independently as a $2\nu$ transition
\cite{factorize}.
The final neutrino fluxes can then 
be written in terms of {\it survival probabilities}
$p$ and $\bar{p}$ of $\nu_e$ and $\bar{\nu}_e$ respectively
(apart from the geometrical factor of $1/R^2$):
\beq
{\small
\left[ \begin{array}{c}
F_e \\ F_{\bar{e}} \\ 4 F_x
\end{array} \right] =
\left[ \begin{array}{ccc}
p & 0 & 1-p \\ 0 & \bar{p} & 1 - \bar{p} \\
1-p & 1 - \bar{p} & 2 + p + \bar{p}
\end{array} \right]
\left[ \begin{array}{c}
F_e^0 \\ F_{\bar{e}}^0 \\  F_x^0
\end{array} \right],}
\label{tr-matrix}
\eeq
where $F_e (F_e^0), F_{\bar{e}} (F_{\bar{e}}^0)$ and
$F_x (F_x^0)$ are the final (initial) fluxes of
$\nu_e, \bar{\nu}_e$ and $\nu_x$ (each of the 
non-electron neutrino or antineutrino species) respectively.


Let $P_H (\bar{P}_H)$ and $P_L (\bar{P}_L)$ be the probabilities that
the neutrinos (antineutrinos) jump to another matter eigenstate in the
resonance layers $H$ and $L$ respectively.
The values of $p$ and $\bar{p}$ are determined by these {\it flip
probabilities} and the mixing matrix elements $|U_{ei}|^2$.

\begin{figure}[htb]
\vspace{9pt}
\epsfig{file=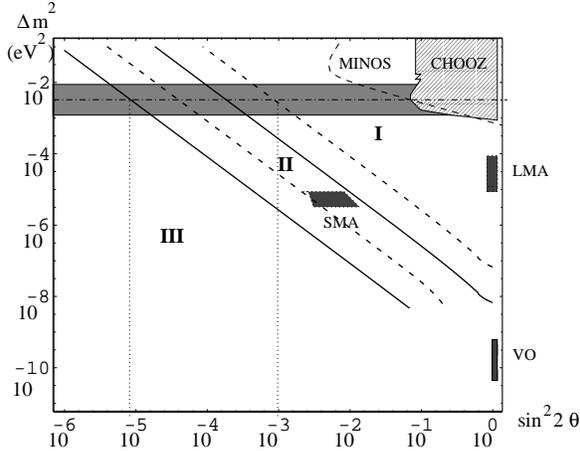,width=3in}
\caption{The contours of equal
flip probability $P_f$.
The solid (dashed) lines denote the contours of flip probability
for a 5 MeV (50 MeV) neutrino: the line on the left stands for
$P_f = 0.9$ (highly non-adiabatic transition)
and the line on the right stands for
$P_f = 0.1$ (adiabatic transition).
}
\label{cont}
\end{figure}

The flip probabilities depend on the density profile
in the resonance layers. Since $\Delta m^2 < 10^{-2}$ eV$^2$
for all transitions, the resonance densities are
$\rho_{res} \lsim 10^4$ g/cc. The resonance layers lie in the outer 
parts of the mantle, where the density profile may be taken as
$r^{-3}$ \cite{profile}. 
Taking $\rho \approx 3.5 \times 10^4$ g/cc $(r/10^9 cm)^{-3}$
as the density profile, we show
in Fig.~\ref{cont} the contours of equal
flip probability $P_f$ 
in the $(\Delta m^2 - \sin^2 2\theta)$ plot
for two different energies on the
borders of the observable spectrum.
The dependence of the contours on the details of the
density profile is very weak \cite{ds}.

The contours of $P_f = 0.1$ and $P_f = 0.9$ divide the plot
into three ``regions''. 
The {\it adiabatic region} (I),
where strong flavour conversions occur,
is the region above the contour with
$P_f = 0.1$.
The {\it non-adiabatic region} (III),
where flavour conversions are almost absent,
lies below
the $P_f = 0.9$ contour.
In the {\it transition region} (II),
the flavour conversions are not complete and
the extent of conversions depends on the 
neutrino energy.

The $H$-resonance lies in the dark horizontal band in 
Fig.~\ref{cont}, which corresponds to the allowed values
of $\Delta m^2_{31}$. 
The features of the final spectra depend strongly on whether the
$H$-resonance lies in the region I or II (See Table~\ref{ppbar}). 
From Fig.~\ref{cont},
the boundary between these regions is at
\beq
\sin^2 2\theta_{e3} \approx 4 |U_{e3}|^2 \sim 10^{-3}~~,
\label{boundary}
\eeq

\section{Observable effects on the final spectra}

The effects of neutrino transitions 
can be observed through the following
features in the neutrino spectra.

(i) The partial or complete disappearance of the $\nu_e$
neutronization peak: the ratio of charged to neutral
current events during the 
neutronization burst is a directly proportional to $p$. 

(ii) The broadening of the spectra: though the initial (pure) spectra
are expected to be ``pinched'' (have a narrower energy distribution
than the Fermi-Dirac one)
\cite{pinched}, the final (mixed) spectra need not be
pinched. This effective broadening \cite{ds}
indicates that the final spectrum is
composite, {\it i.e.} the survival probability ($p$ or $\bar{p}$)
is neither very close to 0 nor to 1.

(iii) The earth matter effects: these can be detected through the
difference in the energy spectra at two detectors, or through 
the distortion of the spectrum \cite{ds} at a single detector.

\begin{table*}[hbt]
\setlength{\tabcolsep}{1.5pc}
\newlength{\digitwidth} \settowidth{\digitwidth}{\rm 0}
\catcode`?=\active \def?{\kern\digitwidth}
\caption{The values of the survival probabilities $p$ and $\bar{p}$
for various scenarios. $[x,y]$ indicates that the value of the 
survival probability lies between $x$ and $y$. In the Earth Matter
Effects columns, a $\surd$ indicates the possibility of
significant earth matter effects.}
\label{ppbar}
\begin{tabular}{rcccc}
\hline
& \multicolumn{2}{c}{Survival Probability} & 
\multicolumn{2}{c}{Earth Matter Effects} \\ 
 & $p$ & $\bar{p}$ & $\nu_e$ & $\bar{\nu}_e$\\
\hline
I ~~SMA ~normal & $|U_{e3}|^2$ & 1 &$\approx 0$ & $\approx 0$ \\
inverted & $P_L$ & $|U_{e3}|^2$ & $\surd$& $\approx 0$\\
 & & & & \\
 LMA ~normal & $|U_{e3}|^2$ & $\cos^2 \theta_\odot$ & $\approx 0$
& $\surd$\\
inverted & $\sin^2 \theta_\odot$ & $|U_{e3}|^2$& $\surd$& $\approx 0$\\ 
 & & & & \\
VO ~normal & $|U_{e3}|^2$ & $[\sin^2 \theta_\odot, 
\cos^2 \theta_\odot]$ & $\approx 0$& $\approx 0$\\
inverted & $[\sin^2 \theta_\odot, 
\cos^2 \theta_\odot]$ &  $|U_{e3}|^2$& $\approx 0$&$\approx 0$\\
\hline
II ~~SMA ~normal & $[|U_{e3}|^2, P_L]$ & 1 & $\surd$& $\approx 0$\\
inverted & $P_L$ & $\bar{P}_H$ & $\surd$& $\approx 0$\\
 & & & & \\
LMA  ~normal & $\sin^2 \theta_\odot P_H$ & $\cos^2 \theta_\odot$ 
& $\surd$& $\surd$\\
inverted & $\sin^2 \theta_\odot$ & $\cos^2 \theta_\odot \bar{P}_H$ 
& $\surd$& $\surd$\\
 & & & & \\
VO ~normal & $[|U_{e3}|^2, \cos^2 \theta_\odot]$ &
$[\sin^2 \theta_\odot, \cos^2 \theta_\odot]$& $\approx 0$&$\approx 0$\\
inverted & $[\sin^2 \theta_\odot, \cos^2 \theta_\odot]$ &
$[\sin^2 \theta_\odot \bar{P}_H, \cos^2 \theta_\odot \bar{P}_H]$ 
&$\approx 0$&$\approx 0$\\
\hline
III ~~SMA ~normal & $P_L$ & 1 &$\surd$&$\approx 0$\\
inverted & $P_L$ & 1 &$\surd$&$\approx 0$\\
 & & & & \\
LMA  ~normal & $\sin^2 \theta_\odot$ & $\cos^2 \theta_\odot$ 
&$\surd$&$\surd$\\
inverted & $\sin^2 \theta_\odot$ & $\cos^2 \theta_\odot$ 
&$\surd$&$\surd$\\
 & & & & \\
VO ~normal & $[\sin^2 \theta_\odot, \cos^2 \theta_\odot]$ &
$[\sin^2 \theta_\odot, \cos^2 \theta_\odot]$ &$\approx 0$&$\approx 0$\\
inverted & $[\sin^2 \theta_\odot, \cos^2 \theta_\odot]$ &
$[\sin^2 \theta_\odot, \cos^2 \theta_\odot]$& $\approx 0$&$\approx 0$\\
\hline
\end{tabular}
\end{table*}

The values of the survival probabilities $p$ and $\bar{p}$
are given in Table~\ref{ppbar}, which illustrate the dependence of the
final neutrino spectra on the ambiguities in the neutrino
mass-flavour spectrum.
From the table, the following observations may be made:


(a)
If the neutronization peak in $\nu_e$ does not disappear
($p > 0.03 \geq |U_{e3}|^2$), and
if the mass hierarchy is known to be the normal one,
the $H$-resonance lies in the region II or III.

(b) If either the final $\nu_e$ or $\bar{\nu}_e$ spectrum is
determined to be hard (almost the original $\nu_x$ spectrum), 
then the $H$-resonance is in region I. This gives
a {\it lower} bound on $|U_{e3}|^2$. Moreover, depending on
which of the spectrum is the hard one, the type of hierarchy
can be determined. A hard $\nu_e$ spectrum implies normal
hierarchy, a hard $\bar{\nu}_e$ spectrum implies inverted
hierarchy.

(c) If both the $\nu_e$ and $\bar{\nu}_e$ can be established
to be composite ({\it e.g.} through the observation of
broadening), the $H$-resonance lies in the region II or III. 

(d)
The observation of any earth matter effects rules out
the scenario with the VO solution.
If the earth matter effects are observed in the
neutrino channels but not in the antineutrino
channels, we either have the inverted mass hierarchy, or
the normal mass hierarchy with the $H$ resonance
in the region II or III.

(e)
The observation of earth matter effects in the antineutrino
channel identifies the LMA solution.
In addition, if the effects are significant
in the neutrino channel also,
the $H$ resonance
can be established to be in the region II or III.

The final neutrino spetra can thus help in resolving the three
kinds of ambiguities. In particular, from the observations
(a), (c), (d), and (e) above, the $H$-resonance can be established
to be in the region II or III, which implies (eq.~\ref{boundary})
an upper bound of
$|U_{e3}|^2 < 10^{-4} - 10^{-3}$.
This is two orders of magnitude better
than the current bound (see CHOOZ in Fig.~\ref{cont}) 
or that expected from the
planned long baseline experiments (see MINOS in Fig.~\ref{cont}).

A galactic supernova may provide us with a sufficient number of events
\cite{gandhi} to enable us to reconstruct the energy spectra
of $\nu_e$ and $\bar{\nu}_e$. 
Here, we have indicated the observations which can be used in principle 
to identify various conversion effects. 
Feasibility studies, involving models for the initial 
spectra and the detector details, are needed in order to
determine what can be achieved in practice.


I would like to thank A. Yu. Smirnov for discussions and insights 
during the collaboration for \cite{ds}. A major part of this work
was performed at The Abdus Salam ICTP, Trieste.

\end{document}